\documentstyle[aps,preprint]{revtex}
\begin{document}
\draft
\preprint{ISSP April No. 2823, 1994, cond-mat/9405028}
\title{
Explicit Solutions of the Bethe Ansatz Equations \\
for Bloch Electrons in a Magnetic Field
}
\author{Yasuhiro Hatsugai$^{1*}$, Mahito Kohmoto$^1$,
and Yong-Shi Wu$^2$}
\address{
$^1$ Institute for Solid State Physics,
 University of Tokyo,
 7-22-1 Roppongi Minato-ku, Tokyo 106, Japan. \\
$^2$ Department of Physics, University of Utah,
 Salt Lake City, Utah 84112, U.S.A.
}

\maketitle
\begin{abstract}
For Bloch electrons in a magnetic field, explicit
solutions are obtained at the center of the spectrum
for the Bethe ansatz equations recently proposed by
Wiegmann and Zabrodin.
When the magnetic flux per plaquette is $1/Q$ where $Q$ is an odd integer,
distribution of the roots is uniform on the unit circle in the
complex plane.
For the semi-classical limit, $ Q\rightarrow\infty$,
the wavefunction obeys the power low and is given
by $|\psi(x)|^2=(2/ \sin \pi x)$
which is critical and unnormalizable.
For the golden mean flux,
the distribution of roots has the exact self-similarity
and the distribution function is nowhere differentiable.
The corresponding wavefunction
also shows a clear self-similar structure.

\end{abstract}

\narrowtext
The Azbel-Hafstadter-Wannier problem of two dimensional
Bloch electron in a magnetic field is an old problem that
often brings new excitements. The equations of motion
can be reduced to one-dimensional ones, which appear
in many different physical contexts, ranging from the
quantum Hall effect\cite{tknn} to quasi-periodic systems
(for a review, see for example ref.\cite{hk}.) Its
topological character has been revealed both for periodic
boundary conditions\cite{tknn} and for systems
with edges\cite{edge_yh}. The interplay of two
intrinsic periods, the period of the lattice and
that of the magnetic flux is essential in this problem.
Previously many studies were done for the cases when
the two periods are commensurable, i.e. the magnetic
flux per plaquette, $\phi$, is rational. But the
incommensurate cases (with an irrational $\phi$) are
more interesting in that the spectrum is
known to have an extremely rich structure
like the Cantor set and to exhibit a multifractal
behaviors.\cite{ht,hk}
Another interesting case is
the weak field limit: When the flux is small, the
semi-classical treatment of the WKB type is
justified. Although some properties of the
wavefunctions are known\cite{tn,aa}, much of the
knowledge we have came from numerical studies for
rational fluxes.\cite{hk} In this letter, we derive
several analytical results for the weak-field limit
and for the incommensurate golden-mean flux, by
following the highly innovative Bethe ansatz approach
recently proposed by Wiegmann and Zabrodin\cite{wz}
using quantum group techniques.

 The Hamiltonian is given by
$
H  =
  \sum_{m,n}
( c_{m+1,n}^\dagger $
$e^{ i \theta^x_{m,n}} c_{m,n}
+
c_{m,n+1}^\dagger e^{ i \theta^y_{m,n}} c_{m,n} + H.c. )
$
where $c_{m,n}$ is the annihilation operator for an electron
at site $(m,n)$.
We first assume the flux per plaquette
$\phi$ is rational: $\phi=P/Q$ with coprime odd integers $P$
and $Q$, and later take appropriate limits of $\phi$.
Let us take the diagonal gauge,
$\theta^x_{m,n}=+\pi\phi(n+m)$, $\theta^y_{m,n}=-\pi\phi(n+m)$.\cite{kh}
Then the Schr\"{o}dinger equation for the
one particle state
$
|\Psi\rangle=\sum_{m,n}
\psi_{m,n} c_{m,n}^\dagger|0\rangle$
is written as
$
i (q^{l+1}+q^{-l}) \Psi_{l+1}-i(q^{l-1}+q^{-l})\Psi_{l-1} = E\Psi_{l}
$
at the mid-band point 
where
$\psi_{m,n}=i^{m+n}\Psi_{m+n}$
and $q=e^{i\pi\phi}=e^{i\pi P/Q}$.
Wiegmann and Zabrodin observed\cite{wz} that
if we write $\Psi_{l}=\Psi(q^l)$ with $\Psi(z)$ a polynomial
of degree $Q-1$, this difference equation coincides
with the eigen-equation of a combination of generators of
the quantum group $U_q(sl_2)$ in the representation
provided by $\Psi(z)$.
Thus the zeros, $\{z_{l}\}$, of the polynomial $\Psi(z)$
must satisfy the Bethe ansatz equations (BAE) \cite{wz}
\begin{equation}
\frac{z_l^2+q}{qz_l^2+1}=-\prod_{m=1}^{Q-1} \frac{qz_l-z_m}{z_l-qz_m},
\ \ l=1,\cdots,Q-1.
  \label{eq:baeq}
\end{equation}
For the zero energy $E=0$, they showed that
$\Psi(z)$ is given by
the so-called continuous q-ultraspherical
polynomial \cite{gr}
as $\Psi(z)= \frac{(q^2;q^2)_n}{(q;q^2)_n}
(-iz)^n P_n (-iz)$, where
$P_n (z) = \sum_{k=0}^n\frac{(q;q^2)_k(q;q^2)_{n-k}}
{(q^2;q^2)_k(q^2;q^2)_{n-k}}z^{n-2k}$, $n=(Q-1)/2$
and $(a;q)_k=\prod_{m=0}^{k-1}(1-aq^m)$.
\cite{another}

In order to understand the properties of the
wavefunctions at $E=0$, the center of the spectrum,
we try to solve the BAE (\ref{eq:baeq}) explicitly. The
polynomial $P_n(z)$ satisfies a difference
equation\cite{exp}
 \begin{equation}
(1-qz^2) P_n(qz)+(q-z^2)P_n(q^{-1}z) = 0.
   \label{eq:dif}
 \end{equation}
Put $z=q^{\pm 1/2}$, then we get $P_n(q^{\pm 3/2})=0$.
So $q^{\pm 3/2}$ are roots.
By iteration, the complete set of roots of the BAE are
given by
\begin{equation}
z_m = i q^{ 2m-1/2}, \ i q^{-2m+1/2}
\ ,\ m=1,\cdots,(Q-1)/2.
  \label{eq:allroots}
\end{equation}
They are all on the unit circle. Let us write the roots
as $z_m=e^{i\theta_m}$ and consider the distribution of
$\theta_m$.

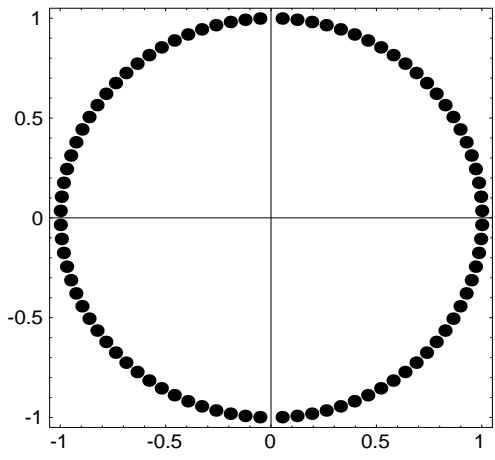
\begin{figure}
\caption{Roots of the Bethe ansatz equation for the special case
$P=1$ and $Q=21$.
 \label{f:unif}}
\end{figure}

Consider first the special case $P=1$.
The roots $\{z_m\}$ distribute uniformly on the unit
circle except near $z=\pm i$.
The roots for $Q=21$ are shown in Fig.~\ref{f:unif}.
In the semi-classical limit $Q\rightarrow \infty$, that is,
$q\rightarrow 1$,
the distribution function
$\rho(\theta)=\lim_{Q\rightarrow\infty}Q\Delta\theta$ is
smooth (constant) where $\Delta\theta=\theta_{m+1}-\theta_{m}$.
A continuous behavior of $\rho(\theta)$ is usually obtained
in the exactly solvable models (the Heisenberg chain,
the Hubbard chain, etc.) in which $\rho(\theta)$ is
determined by an integral equation.

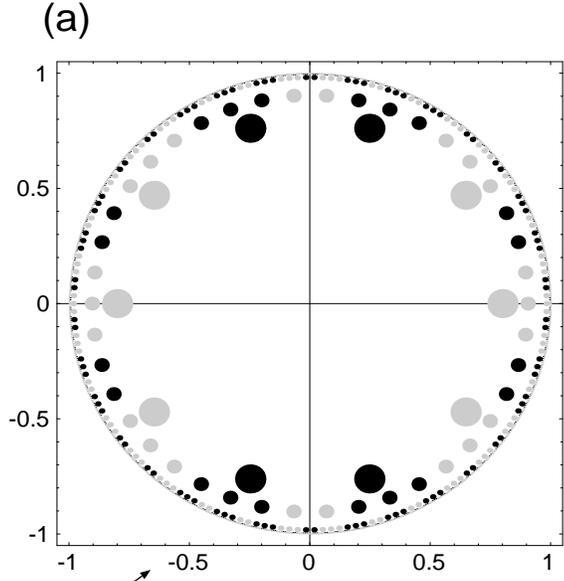
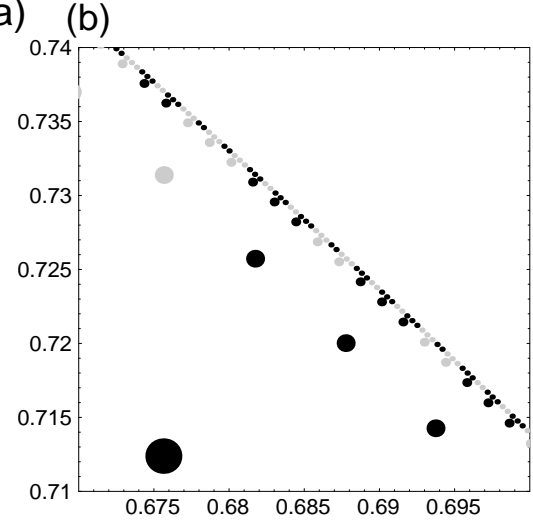
\begin{figure}
\caption{Roots of the Bethe ansatz equation for the ratio of the
Fibonacci numbers.
\hfil\break
$\phi_k=P_k/Q_k=3/5,$$ 13/21,$$ 55/89,$$ 233/377,$$ 987/1597,$$ 4181/6765$
(a) in the whole complex plane and (b) enlarged figure.
 In each case,
the roots are always on the unit circle. We scaled the radii to
show the branching rule clearly.
 \label{f:incom}}
\end{figure}

When the flux is irrational, the situation is quite different.
Take the flux $\phi=1/\tau=(\sqrt{5}-1)/2$, where $\tau$ is
the golden mean. To reach this flux, we consider a sequence of
rational fluxes $\phi_k=P_k/Q_k$, where $Q_k=F_{3k+1},
\ P_k=F_{3k}$ and $F_k$ is a Fibonacci number
defined by $F_{k+1}=F_{k}+F_{k-1},\ F_1=1,\ F_0=1$.
In this case, the two types of roots in Eq.~(\ref{eq:allroots})
are nested. To gain an insight into the distribution of roots,
it is helpful to consider the pseudo roots which are defined
also by Eq.~(\ref{eq:allroots}) but with the range of $m$
modified to $m=-(Q-1)/2,\cdots,0$. In Figs.~\ref{f:incom}
we show the distributions of the roots (black points) and
pseudo roots (gray points) for several $\phi_k,(\ k=1,2,3,4,5,6)$.
Here the radius of the unit circle has been scaled so as
to show all the cases at once. These figures clearly show that
there is a branching rule for the true roots (denoted by $A$)
and pseudo roots (by $B$) as follows:
\begin{eqnarray}
A^3 &\rightarrow & A^3B^2A^3B^2A^3 \nonumber \\
A^2 &\rightarrow & A^3B^2A^3 \nonumber \\
B^3 &\rightarrow & B^3A^2B^3A^2B^3 \nonumber \\
B^2 &\rightarrow & B^3A^2B^3.
  \label{eq:rule}
\end{eqnarray}
The initial condition is $B^3A^2B^3A^2$ (cyclic).
In the $k$-th stage of the sequence,
the number of clusters of the true roots
$A^3$ and $A^2$ are $Q_{k-1}-1$ and
$P_{k-1}+1$ respectively. This branching rule gives
rise to a self-similar structure for the
distribution $\rho(\theta)$ in the limit $k \to\infty$.
To characterize the distribution, let us define the
generation of a root. According to the branching
rule (\ref{eq:rule}), each true (pseudo) root branches
into a cluster of 3 new true (pseudo) roots,
each of which in a sense
has a parent. At the same time, between these
clusters of new pseudo (true) roots, there is a pair of
new-born true (pseudo) roots which have no parent.
We assign the generation number to a root so that it
is $1$ when the root does not have a parent,
otherwise it is one plus that of its parent.
Let us denote the number of true (pseudo) roots
in the $k$-th stage with generation $g$
by $n_A(g,k)\  (n_B(g,k))\ (g=1,\cdots,k)$.
(Then in the special case $P=1$,
$n_A(g,k)=n_B(g,k)=\delta_{kg}$.)
In the present case, we get a recursion formula
by the branching rule as
$n_A(g,k)=3 n_A(g-1,k-1), \ g=2, \cdots , k ,\ \
n_A(1,k)= 2 (P_{k-1} +1)$.
Thus $n_A(g,k)=2\cdot 3^{g-1} (P_{k-g}+1)$ and
$n_B(g,k)=2\cdot 3^{g-1} (P_{k-g}-1)$.
The above considerations exemplify the difference
of the distributions between the semi-classical
limit and the incommensurate case. The distribution
of the roots has a self-similar structure and
the function $\rho(\theta)$ is nowhere differentiable
in the incommensurate limit, while that for the
semi-classical limit is smooth. We believe this is
characteristic to the incommensurate case.

Another way to characterize the distribution  is
to map to the dual (reciprocal) space.
This can be done for arbitrary $P$ and $Q$.
We lift the $\theta_m$ to the real axis periodically.
On the real axis , the true and pseudo roots occupy
a lattice of points $\{j/2Q\ |\ j {\rm : integer} \}$ with spacing $1/2Q$.
Thus we perform the Fourier transform by
$S_Q(p)=\sum_{j=-\infty}^\infty e^{i p j} \tilde S_Q(j)$ where
$\tilde S_Q(j)$ is the so-called defining function:
$\tilde S_Q(j)=1$ if there is a true root at $j/2Q$,
otherwise $\tilde S_Q(j)=0$. Then
\begin{equation}
 S_Q(p)=\frac{\pi}{Q}
\sum_{r=0}^{Q-1}s_Q^r\,{\delta(p-p_r)}, \ (0\leq p<2 \pi ),
\label{eq:rootfrt}
\end{equation}
where $s_Q^0=(Q-1)$,
$s_Q^r=(-)^{r+1}[\cos(\frac 1 2 P p_r)]^{-1}$
$(r=1,\cdots Q-1)$, and $p_r=2\pi r /Q$.
In the semi-classical limit,
$|S_Q(p)|^2$ is well defined and behaves smoothly.
On the other hand, $|S_Q(p)|^2$ is not even differentiable
in the incommensurate limit since $P\rightarrow\infty$.
Also it can be shown that the original defining
function is given by
$\tilde S_Q(j) = (\frac 1 {2Q}) \sum_{n=1}^{Q-1}
[ 1-(-)^n{\cos(2\pi j n/Q)}/{\cos(\pi P n/Q)}]$.

Next let us consider the wavefunction.
Using the explicit roots we found above,
the wave function at site $j$ can be written
in a compact factorized form
 \begin{equation}
\Psi_j= (-q)^{-j}(i q^{-j+3/2};q^2)_{(Q-1)/2}
(iq^{-j-3/2};q^{-2})_{(Q-1)/2}.
   \label{eq:wfna}
\end{equation}
It is convenient to shift the site, $\bar j=j+j_0$, by
an amount $j_0$
where $j_0$ is determined by $Pj_0=(Q-P)/2 \  ({\rm mod}\ 2Q\ )$.
Then $\Psi_{\bar j}=0$ at $\bar j=2m, -2m+1$
($m=1,\cdots,(Q-1)/2$), and the wavefunction is
nonzero only at $\bar j=1,3,\cdots,Q$ and
$Q+1,Q+3,\cdots 2Q$. Using Eq.~(\ref{eq:wfna}),
the amplitude of the wavefunction is
calculated as
\begin{equation}
|\Psi_{\bar j}|^2 = \prod_{m=1}^{(Q-1)/2}
| 4  \sin {\frac{\pi(2m-\bar j)P}{2Q}}
 \sin {\frac{\pi(2m+\bar j-1)P}{2Q}} |^2.
  \label{eq:amp}
\end{equation}

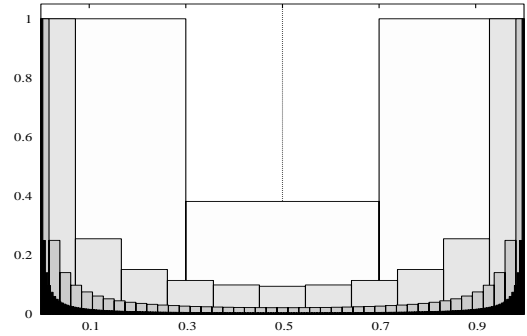
\begin{figure}
\caption{Squared amplitudes
for the wavefunctions: the special cases $P=1$ and
$Q=5,21,89,377$. The wavefunctions are normalized by the peak heights.
 \label{f:wfnsemi}}
\end{figure}

First let us present the exact results explicitly
for the special case $P=1$.
A direct and careful calculation leads to a very simple
result for the amplitude
\begin{equation}
|\psi (x)|^2 = \frac{2}{ \sin(\pi x)},\ \ \ (0<x<1),
  \label{eq:wfn}
\end{equation}
up to a constant factor, where
$\psi(x)=\Psi_{\bar j},\ x=(2\bar j -1)/2Q$,
and the semi-classical limit $Q\rightarrow
\infty$ is taken. Thus the squared amplitude of
the wavefunction is given by the
inverse chord distance in the semi-classical limit.
The recursion relation
$|\Psi_{\bar j}|^2=|\Psi_{\bar j-2}|^2
\sin^2(\pi(j-2)/Q)/\sin^2(\pi(j-1)/Q)$
obtained from Eq.~(\ref{eq:amp}),
has played a key role. \cite{ini}
For a finite $Q$,
a correction factor appears only near the edges
($x=0$ and $1$).
For example, near the edge $x\approx 0$, the finite size
correction is given by
$|\psi (x_{2l+1})|^2   =   C(l) (2/ \sin\pi x_{2l+1})$,
with $ C(l) =  \pi(l+1/4)\prod_{k=1}^l(1-1/2k)^2$,
where $C(0)=\pi/4=0.78539...,\ C(1)=5\pi/16=0.98174...,
\ C(2)=81\pi/256=0.99402...,\cdots$.
So the finite-size correction factor $C(l)$ converges
to unity very rapidly and the Eq.~(\ref{eq:wfn}) is
quite accurate even at small $Q$. The norm of the
wavefunction is $\log Q+{\rm const.}$ and
unnormalizable which is characteristic to a critical wavefunction.
In Fig.~\ref{f:wfnsemi}, the amplitudes of the
analytic wavefunctions, normalized by the peak height,
are shown for several values of $Q$.

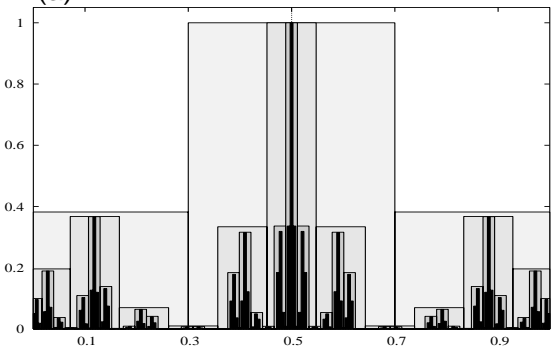
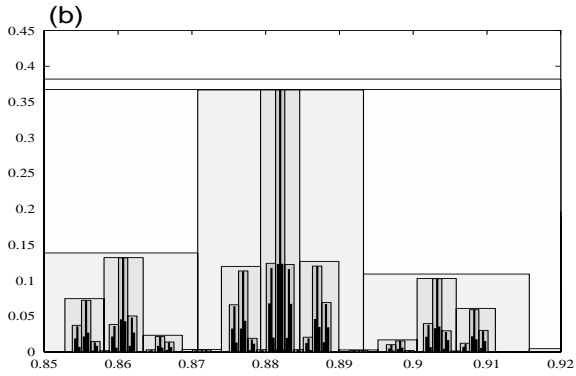
\begin{figure}
\caption{Amplitude of the wavefunctions: for the ratios of
successive Fibonacci numbers:
(a) whole region ($\phi=3/5,$ $ 13/21,$ $ 55/89 $ )
and
(b) enlarged plot in the region near 0.89.
($\phi=P_k/Q_k=3/5,$ $ 13/21,$ $ 55/89,$ $ 233/377,$ $ 987/1597,$
$ 4181/6765$)
The wavefunction with larger value of $Q$ is shaded darker.
The wavefunctions are normalized by the peak height.
 \label{f:wfninc}}
\end{figure}

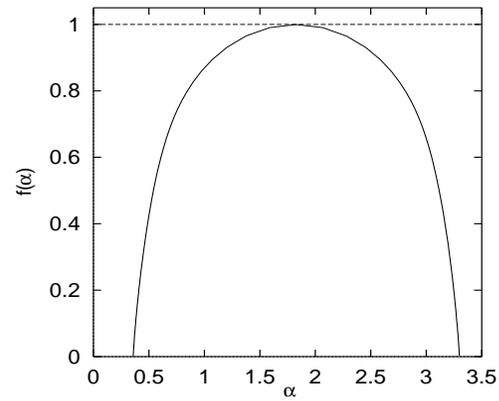
\begin{figure}
\caption{The plot of $f(\alpha)$ for the wavefunction in the golden mean
case.
 \label{f:mf}}
\end{figure}

Next let us discuss the case with golden-mean flux.
We plot the analytic results  Eq.~(\ref{eq:amp})
in Fig.~\ref{f:wfninc} for a sequence of rational
fluxes converging to $1/\tau$. One can easily recognize
the self-similar behavior of the wave function.
Each peak branches into three peaks in the next stage.
Presumably these are the reflection of the
self-similar distribution of the roots, Eq.~(\ref{eq:rule}).
 The multifractal analysis\cite{hk}
 done numerically gives a smooth $f(\alpha)$
shown in Fig.~\ref{f:mf}.
This clearly shows that this wavefunction is multifractal and critical.
We note the striking resemblance of these wavefunctions to that
of the $1d$ quasicrystal Fibonacci lattice at the center of
the spectrum.\cite{kb}
An analytical derivation of $f(\alpha)$ is under progress.
The latter was obtained exactly by a different technique and
$f(\alpha)$ is obtained analytically.\cite{fkt}

In conclusion, we have found analytical solutions
to the BAE (\ref{eq:baeq}) that describes
the Bloch state in a magnetic field with zero energy.
The flux per plaquette is $\phi=P/Q$  with
coprime odd integers $P$ and $Q$.
All roots are on the unit circle, and the defining
function for the roots is explicitly derived.

When $P=1$, the roots distribute uniformly
for any odd $Q$. In the semi-classical
limit $Q\rightarrow\infty$, the density function of
roots is smooth. When the flux $\phi$ are ratios
of the successive Fibonacci numbers which converge
to the golden mean, we found the branching rule
for the roots, which makes the density function
$\rho$ of the roots exactly self-similar and
nowhere differentiable.

We are able to get explicit wavefunctions too.
In the semi-classical case, a compact expression
is derived for the critical wavefunction,
which turns out not normalizable. The squared
amplitude, $|\psi(x)|^2$, is given by the
inverse chord distance, $2/\sin(\pi x)$. For
the golden mean flux, the wavefunction also
has a clear self-similar branching structure.

Finally we remark that the solutions (\ref{eq:allroots})
to the BAE has a relatively simple structure, and
positions of the roots are not affected by a finite
$Q$ correction.
These facts together
with the simple form (\ref{eq:wfn}) of the amplitude
of the wavefunctions suggest that the symmetry of the
problem, at least at the center of the spectrum,
is higher than originally expected. It might just be
the quantum group $U_q(sl_2)$, though a clear-cut proof
is still lacking. The situation would be similar to
the Haldane-Shastry model where the symmetry is known
to be higher than what is naively expected.\cite{hl,st}

\begin{references}

\bibitem[*]{electric-address}
electronic mail address: hatsugai@tansei.cc.u-tokyo.ac.jp

\bibitem{tknn}D. J. Thouless, M. Kohmoto, P. Nightingale, and M. den Nijs,
\ Phys.\ Rev.\ Lett.\  {\bf 49}, 405 (1982).

\bibitem{hk} H. Hiramoto and M. Kohmoto, \ Int.\ J.\ Mod.\ Phys.\
B\ {\bf 6}, 281 (1992).

\bibitem{edge_yh}Y. Hatsugai, \ Phys. \ Rev. \ B {\bf 48}, 11851 (1993),
 \ Phys.\ Rev.\ Lett.\  {\bf 71}, 3697 (1993).

\bibitem{ht}D. R. Hofstadter, \ Phys. \ Rev. \ B {\bf 14}, 2239 (1976).

\bibitem{tn}D. J. Thouless and Q. Niu
\ J.\ Phys. \ A\  {\bf 16}, 1911 (1983).

\bibitem{aa}S. Aubry and G. Andr\'{e}, \ Ann.\ Israel\ Phys.\
Soc., {\bf 3}, 131 (1980).

\bibitem{wz}P. B. Wiegmann and A. V. Zabrodin,
 \ Phys.\ Rev.\ Lett.\  {\bf 72}, 1890 (1994). See also Ref. 8.

\bibitem{fd}L. D. Fadeev and R. M. Kashaev, University of
Helsinky report, preprint HU-TFT-93-63.

\bibitem{kh} M. Kohmoto and Y. Hatsugai, \ Phys. \ Rev. \ B {\bf 41},
9527 (1990).


\bibitem{gr} This $P_n(z)$ is obtained from the continuous q-ultraspherical
polynomials as $P_n(z)=C_n((z+z^{-1})/2);q|q^2)$. See p169
in G. Gasper and M. Rahman,
{\it Basic Hypergeometric Series}, Cambridge Univ. Press 1990.

\bibitem{another}
Our 1d difference equation ( in the diagonal gauge) at $E=0$
has another solution which is given by
$
\tilde \Psi_l=\tilde \Psi(q^l)$ where
$\tilde \Psi(z)=  \frac{(q^2;q^2)_n}{(q;q^2)_n}$
$ (-iz^{-1})^{n+1} P_n(iz^{-1})$
which has a pole at $z=0$.
This solution does not satisfy the pole free condition,
and separate consideration is needed.
However, since $\{z_m\}=\{z_m^*\}$ as shown later,
we have $|\Psi_l|=|\tilde\Psi_l|$. Thus we only consider
$\Psi_l$ from now on.


\bibitem{exp}
This can be derived directly from the definition.

\bibitem{ini}
$ |\Psi_{\bar j=1}|^2=Q $ for arbitrary $\phi=P/Q$.

\bibitem{kb} M. Kohmoto and J. R. Banavar,
\ Phys.\ Rev.\ B\  {\bf 34}, 563 (1986).

\bibitem{fkt} T. Fujiwara, M. Kohmoto, T. Tokihiro,
\ Phys.\ Rev.\ B\  {\bf 40}, 7413 (1989).

\bibitem{hl} F. D. M. Haldane,
\ Phys.\ Rev.\ Lett.\  {\bf 60}, 635 (1988),
F. D. M. Haldane, Z. N. C. Ha, J. C. Talstra, D. Bernard, and V. Pasquier,
\ Phys.\ Rev.\ Lett.\  {\bf 69}, 2021 (1992),

\bibitem{st} B. S. Shastry,
\ Phys.\ Rev.\ Lett.\  {\bf 60}, 639 (1988).


\end{references}
\end{document}